\documentclass[12pt]{article}
\usepackage[utf8]{inputenc}
\usepackage{authblk}
\usepackage{lipsum}
\usepackage{fancyhdr}
\usepackage{geometry}
\usepackage{titlesec}
\usepackage{setspace}
\usepackage{amssymb}
\usepackage{graphicx}
\usepackage{subcaption}
\usepackage[backend=biber,style=numeric,sorting=none]{biblatex}
\DeclareFieldFormat[article]{title}{\textbf{#1}}

\DeclareFieldFormat[article]{title}{\textbf{#1}}
\DeclareFieldFormat[inproceedings]{title}{\textbf{#1}}  
\DeclareFieldFormat[inbook]{title}{\textbf{#1}}        
\DeclareFieldFormat[book]{title}{\textbf{#1}}         
\DeclareFieldFormat{journaltitle}{#1}  
\renewbibmacro{in:}{}  

\defbibheading{bibliography}{\section*{\textbf{References}}}

\title{\textbf{Techniques of Artificial Intelligence Applied to Near-Infrared Spectra}}

\author{
  Aminata Sow\textsuperscript{1,*},
  Tidiane Diallo\textsuperscript{2}
}

\date{}

\begin{document}

\maketitle

\begin{center}
\textsuperscript{1}Department of Physics, Faculty of Sciences and Techniques (FST),\\
University of Sciences, Techniques and Technologies of Bamako (USTTB), Mali \\
\texttt{aminasow100@gmail.com}, \texttt{aminata.sow@usttb.edu.ml} \\[1em]

\textsuperscript{2}Department of Drug Sciences, Faculty of Pharmacy,\\
University of Sciences, Techniques and Technologies of Bamako (USTTB), Mali \\
\texttt{tidiallo2017@gmail.com} \\[1em]

\textsuperscript{*}Corresponding author
\end{center}

\maketitle

\begin{abstract}
This article explores the application of various artificial intelligence techniques to the analysis of near-infrared (NIR) spectra of paracetamol, within the spectral range of 900~nm to 1800~nm. The main objective is to evaluate the performance of several dimensionality reduction algorithms—Principal Component Analysis (PCA), Kernel PCA (KPCA), Sparse Kernel PCA, t-Distributed Stochastic Neighbor Embedding (t-SNE), and Uniform Manifold Approximation and Projection (UMAP)—in modeling and interpreting spectral features. These techniques, derived from data science and machine learning, are evaluated for their ability to simplify analysis and enhance the visualization of NIR spectra in pharmaceutical applications.
\end{abstract}

\textbf{Keywords:} Near-Infrared (NIR) Spectroscopy; Dimensionality Reduction; PCA; KPCA; t-SNE; UMAP; Pharmaceutical Analysis.

\newpage

\section{Introduction}

Near-infrared (NIR) spectroscopy, particularly when combined with chemometric techniques, has proven to be a powerful analytical tool across a wide range of scientific and industrial applications. Its non-destructive nature, rapid analysis time, and ability to handle complex mixtures make it especially valuable in the pharmaceutical, agricultural, food, and environmental sectors.

In the pharmaceutical domain, NIR spectroscopy has been widely used for the detection and classification of chemical substances with high accuracy. For example, Risoluti et al.~\cite{risoluti_early_2016} addressed the early detection of new psychoactive substances, such as cannabinoids and phenethylamines, by applying NIR spectroscopy in combination with chemometric techniques—an approach crucial for mitigating their growing public health impact. Similarly, Kos et al.~\cite{kos_unveiling_2025} provide a comprehensive overview of recent advances in NIR spectroscopy for biomedical and pharmaceutical applications. Beyond pharmaceuticals, NIR spectroscopy is also employed in the food industry for quality control and compositional analysis~\cite{fodor_role_2024, squeo_considerations_2024}, and in agriculture and environmental science to study soil composition and monitor ecosystem health~\cite{bellon-maurel_critical_2010, carra_near-infrared_2019}.

In parallel with these advances, machine learning techniques have become increasingly important in the interpretation and analysis of spectral data. Supervised learning methods, such as regression and classification, are commonly applied for predictive modeling. In contrast, unsupervised learning techniques—particularly clustering and dimensionality reduction—are employed to uncover hidden patterns, group structures, and meaningful representations in high-dimensional datasets.

Among the most widely used dimensionality reduction techniques is Principal Component Analysis (PCA)~\cite{pearson1901}, which projects data into a lower-dimensional space by maximizing variance along orthogonal directions. However, PCA is limited to capturing only linear relationships. Kernel PCA (KPCA)~\cite{scholkopfkpca} addresses this limitation by applying a kernel function to project the data into a high-dimensional feature space, enabling the extraction of non-linear structures. Further extending this approach, Sparse Kernel PCA (SKPCA)~\cite{mika1999skpca} introduces sparsity constraints to improve computational efficiency and interpretability. In addition, manifold learning techniques such as t-distributed Stochastic Neighbor Embedding (t-SNE)~\cite{maaten2008tsne} and Uniform Manifold Approximation and Projection (UMAP)~\cite{mcinnes2018umap} have gained popularity for visualizing high-dimensional data, particularly for their ability to preserve local and global structures.

These developments underscore the importance of combining NIR spectroscopy with data-driven approaches to extract meaningful insights from complex datasets. In our previous work~\cite{sow2022-1, Sow2022}, we applied chemometric algorithms to analyze the NIR spectra of paracetamol, uncovering key structural and spectral characteristics. Building on these findings, the present study explores the application of advanced unsupervised machine learning techniques—particularly dimensionality reduction—to further investigate the spectral behavior of paracetamol and reveal underlying patterns within the dataset.

The remainder of this manuscript is organized as follows: Section~\ref{sec:methods} reviews the dimensionality reduction algorithms employed in this study. Section~\ref{sec:results} presents and discusses the results obtained from applying these techniques to the NIR spectra of paracetamol. Finally, Section~\ref{sec:conclusions} offers concluding remarks of this investigation.

\section{Dimensionality Reduction Techniques}
\label{sec:methods}

Measurement using near-infrared (NIR) spectroscopy and similar instruments often yields high-dimensional spectral data. Such data typically contain redundant or highly correlated features, which can obscure the underlying structure and negatively impact the performance of machine learning algorithms—a challenge commonly referred to as the curse of dimensionality \cite{awotunde22, manley2014nir, binson2024}. To mitigate this issue, dimensionality reduction techniques are employed to transform high-dimensional data into a lower-dimensional space while retaining the most relevant and informative features. These techniques are broadly classified into linear and non-linear methods, depending on how they capture the intrinsic structure of the data. A comprehensive overview of these approaches is provided in Ref.~\cite{sorzano2014}.

\subsection{Principal Component Analysis (PCA)}

Principal Component Analysis (PCA)~\cite{pearson1901} is one of the oldest and most widely used linear dimensionality reduction techniques. The mathematical foundation of PCA is thoroughly explained in Ref.~\cite{jolliffe2002pca}. In this algorithm, the data is projected onto a new set of orthogonal axes, known as \textit{principal components}, which are ordered by the amount of variance they capture from the original data. The covariance matrix plays a central role in identifying these components. Typically, the first few principal components capture the majority of the variance, enabling effective data compression and visualization while preserving most of the important information contained in the original data.

In the context of this work, PCA is applied to near-infrared (NIR) spectral data of paracetamol to reduce dimensionality and highlight the most informative variance in the dataset. This step is crucial not only for efficient data representation but also for enhancing interpretability by removing noise and redundant features from the spectral measurements.

\subsection{Kernel PCA (KPCA)}

While PCA is inherently a linear technique, it may not effectively capture complex non-linear relationships present in the data. To address this limitation, Kernel PCA (KPCA) was introduced by Schölkopf et al.~\cite{scholkopfkpca}. KPCA extends the capabilities of PCA by mapping the input data into a high-dimensional feature space using a kernel function—such as the Gaussian (RBF) kernel, which is used in our application, or a polynomial kernel. In this transformed feature space, linear PCA is performed, allowing the extraction of non-linear structures that standard PCA cannot detect.

In this manuscript, KPCA is employed to analyze the non-linear patterns in the near-infrared (NIR) spectral data of paracetamol. By capturing more complex relationships between spectral features, KPCA enhances the ability to classify spectral data more effectively.

\subsection{t-Distributed Stochastic Neighbor Embedding (t-SNE)}

t-distributed Stochastic Neighbor Embedding (t-SNE) was introduced by van der Maaten and Hinton~\cite{maaten2008tsne}. The t-SNE algorithm is a non-linear technique designed for visualizing high-dimensional data in two or three dimensions. It models pairwise similarities between data points using probability distributions and minimizes the Kullback–Leibler divergence between the distributions in the original and reduced spaces. t-SNE is particularly effective at preserving local structures and revealing clusters in the data, but it is sensitive to parameters such as perplexity and learning rate.

\subsection{Uniform Manifold Approximation \& Projection (UMAP)}

UMAP was introduced by McInnes et al.~\cite{mcinnes2018umap}. UMAP is a more recent non-linear dimensionality reduction technique that is based on manifold learning and topological data analysis. It constructs a high-dimensional graph representation of the data and optimizes a low-dimensional graph to be structurally similar. Compared to t-SNE, UMAP tends to preserve both local and global structures better and is computationally more efficient for large datasets.

\subsection{Sparse Kernel PCA}

Sparse Kernel PCA (SKPCA) was first proposed by Mika et al.~\cite{mika1999skpca} as an extension of Kernel PCA (KPCA) that incorporates sparsity constraints. By reducing the number of components or support vectors involved in the transformation, SKPCA enhances interpretability and significantly lowers computational cost—especially beneficial when working with large or high-dimensional datasets.

Each of the dimensionality reduction techniques discussed—PCA, KPCA, SKPCA, t-SNE, and UMAP—offers distinct strengths and trade-offs in terms of computational efficiency, interpretability, and preservation of data structure. In the following section, we apply these methods to the NIR spectra of paracetamol and compare their effectiveness in capturing meaningful spectral patterns and relationships.

\section{Results and Discussion}\label{sec:results}

In this section, we present the results of applying various dimensionality reduction techniques to the near-infrared (NIR) spectra of paracetamol. Our objective is to evaluate each method’s ability to uncover relevant structural patterns, reduce noise, and facilitate effective visualization of the spectral data. The techniques evaluated include Principal Component Analysis (PCA), Kernel PCA (KPCA), Sparse Kernel PCA (SKPCA), t-Distributed Stochastic Neighbor Embedding (t-SNE), and Uniform Manifold Approximation and Projection (UMAP).

As a preliminary step, the samples were divided into two classes based on their content values. Class 1 consists of samples with content greater than 95 and less than 1015, while Class 2 contains the remaining samples.
 
\subsection{Data Preprocessing}

Prior to applying dimensionality reduction techniques, the spectral data were preprocessed to enhance signal quality and reduce irrelevant variability. Standard preprocessing steps included standard normal variate (SNV), detrending, and, where necessary, multiplicative scatter correction (MSC).

\begin{figure}[htbp]
    \centering
    
    \begin{subfigure}[b]{0.45\textwidth}
        \centering
        \includegraphics[width=\linewidth]{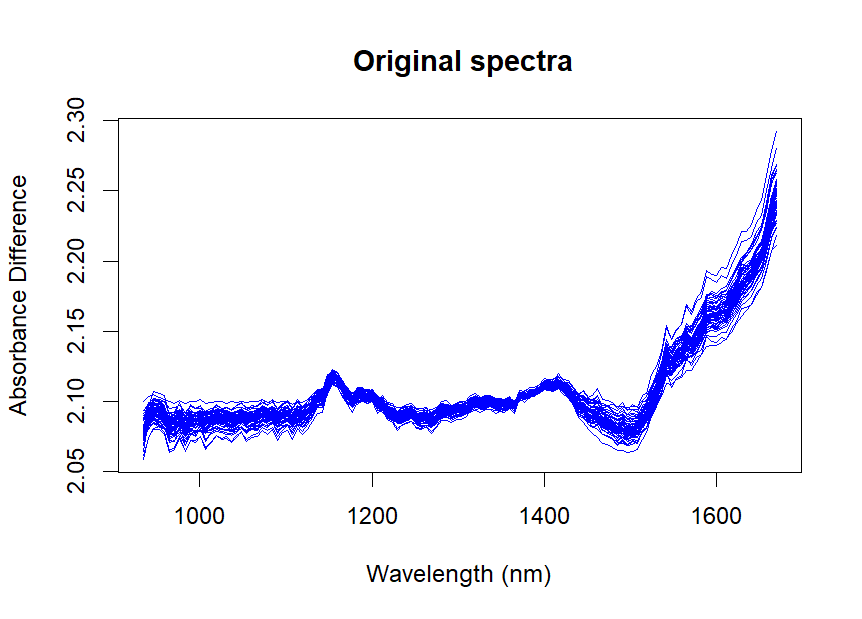}
        \caption{Original spectra}
        \label{fig:original}
    \end{subfigure}
    \hfill
    \begin{subfigure}[b]{0.45\textwidth}
        \centering
        \includegraphics[width=\linewidth]{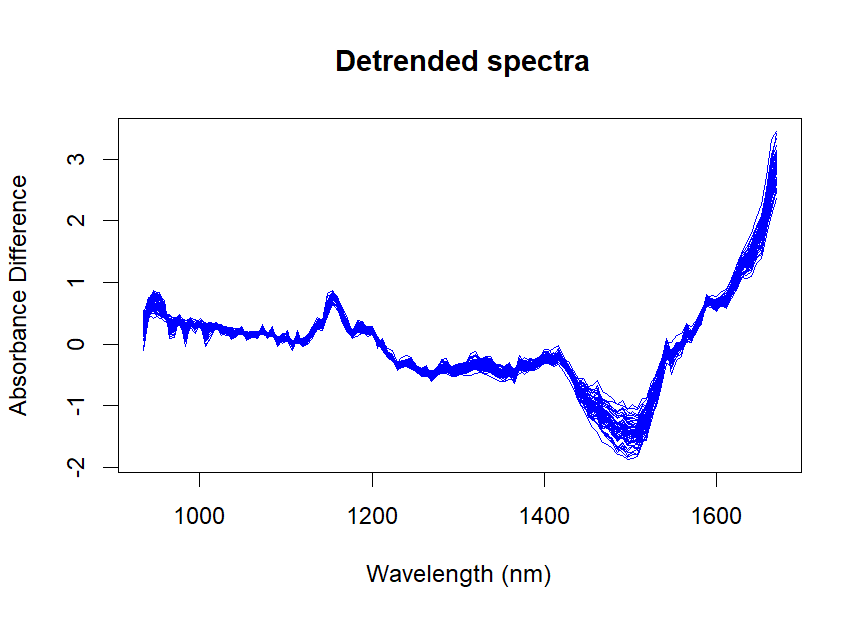}
        \caption{Detrended spectra}
        \label{fig:detrended}
    \end{subfigure}

    \vspace{0.5cm} 

    \begin{subfigure}[b]{0.45\textwidth}
        \centering
        \includegraphics[width=\linewidth]{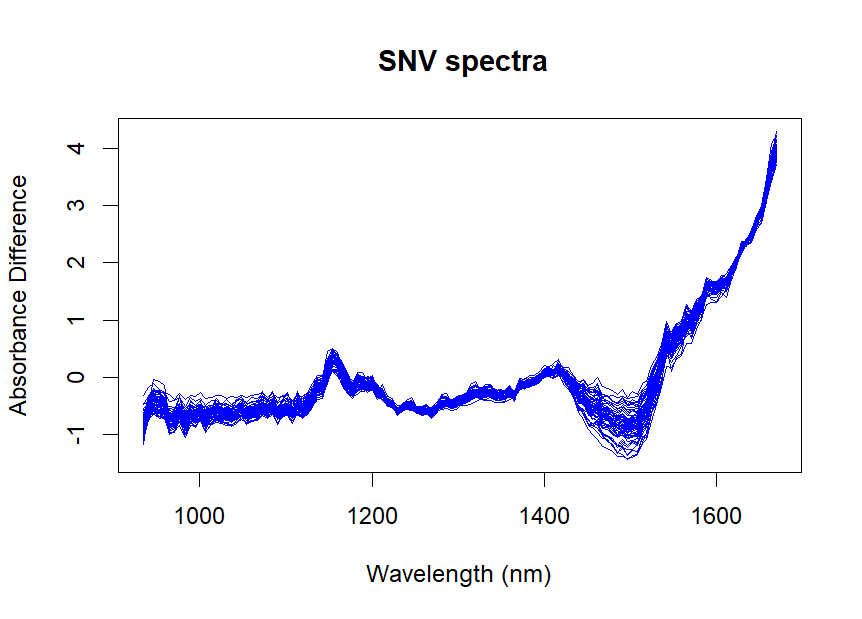}
        \caption{SNV-corrected spectra}
        \label{fig:snv}
    \end{subfigure}
    \hfill
    \begin{subfigure}[b]{0.45\textwidth}
        \centering
        \includegraphics[width=\linewidth]{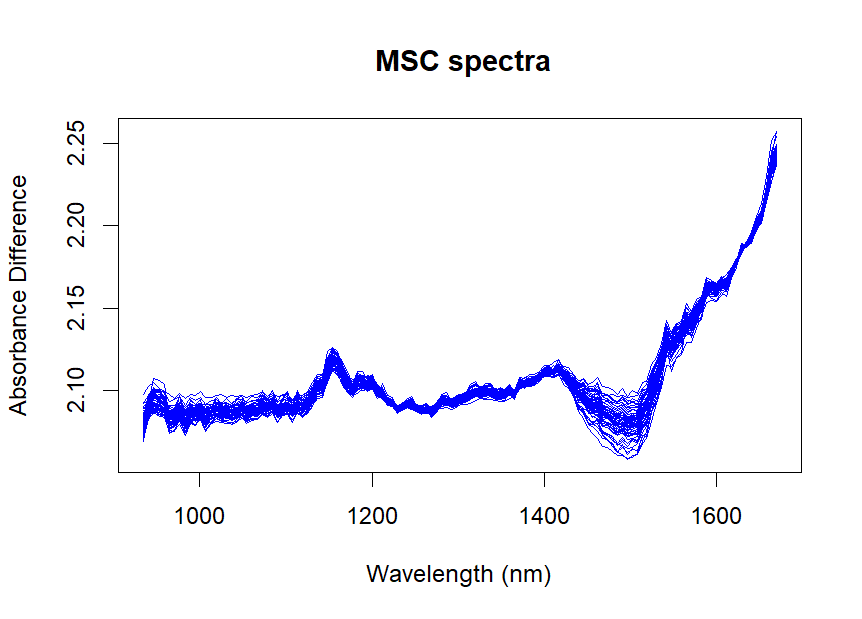}
        \caption{MSC-corrected spectra}
        \label{fig:msc}
    \end{subfigure}
    
    \caption{Spectral data before and after preprocessing using detrending, standard normal variate (SNV), and multiplicative scatter correction (MSC).}
    \label{fig:preprocessing}
\end{figure}

Figure~\ref{fig:preprocessing} presents the original spectra alongside the preprocessed spectra using the aforementioned techniques.

\subsection{Visualization of Reduced Representations}

\begin{figure}[htbp]
    \centering
    \begin{subfigure}[b]{0.45\textwidth}
        \centering
        \includegraphics[width=\textwidth]{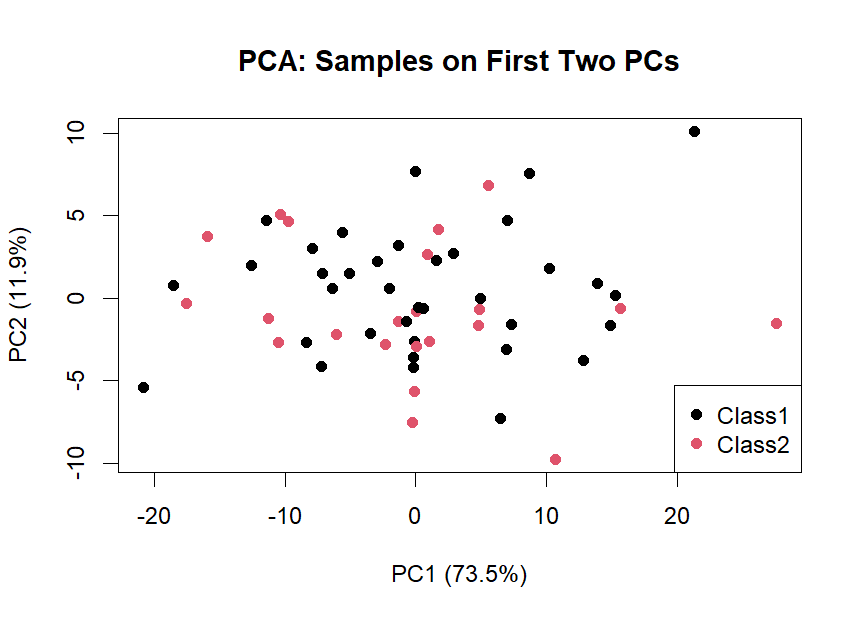}
        \caption{PCA}
        \label{fig:pca}
    \end{subfigure}
    \hfill
    \begin{subfigure}[b]{0.45\textwidth}
        \centering
        \includegraphics[width=\textwidth]{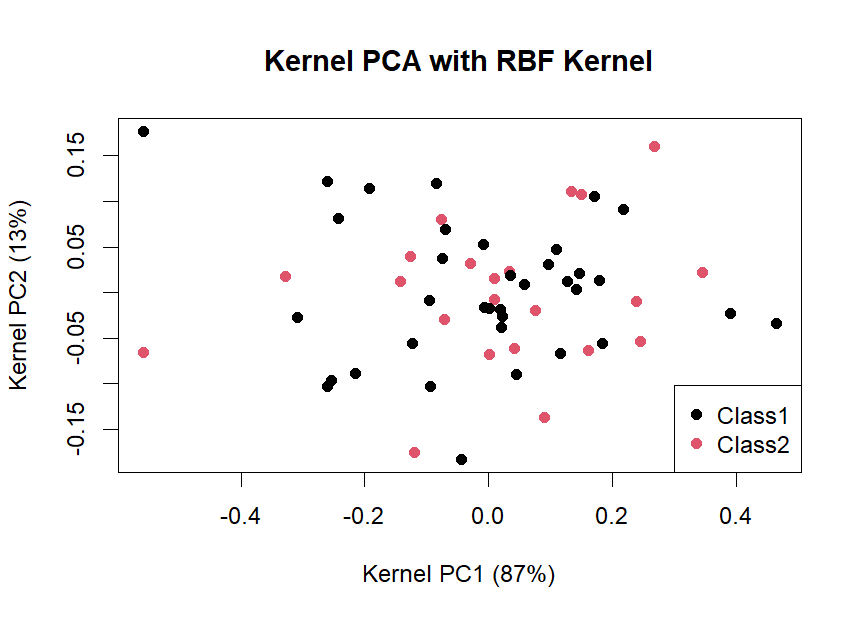}
        \caption{Kernel PCA}
        \label{fig:kpca}
    \end{subfigure} 
    \caption{Embeddings of NIR spectral data using linear and kernel-based PCA techniques.}
    \label{fig:embeddings1}
\end{figure}

\begin{figure}[htbp]
    \centering
    \begin{subfigure}[b]{0.45\textwidth}
        \centering
        \includegraphics[width=\textwidth]{kpcaok.png}
        \caption{Kernal PCA}
        \label{fig:pca}
    \end{subfigure}
    \hfill
    \begin{subfigure}[b]{0.45\textwidth}
        \centering
        \includegraphics[width=\textwidth]{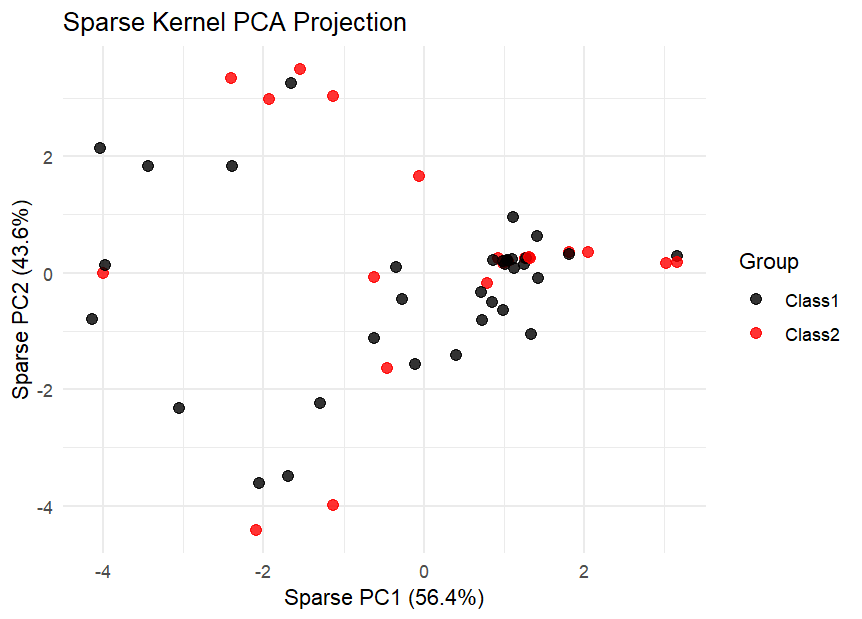}
        \caption{Sparse Kernel PCA}
        \label{fig:kpca}
    \end{subfigure} 
    \caption{Embeddings of NIR spectral data using  sparse and kernel-based PCA techniques.}
    \label{fig:embeddings2}
\end{figure}

\begin{figure}[htbp]
    \centering
    \begin{subfigure}[b]{0.45\textwidth}
        \centering
        \includegraphics[width=\textwidth]{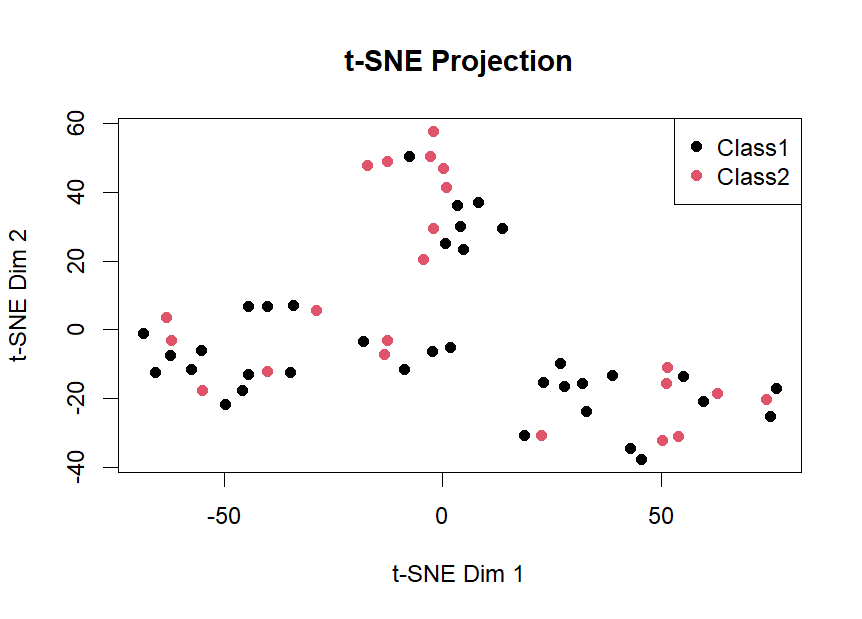}
        \caption{t-SNE}
        \label{fig:tsne}
    \end{subfigure}
    \hfill
    \begin{subfigure}[b]{0.45\textwidth}
        \centering
        \includegraphics[width=\textwidth]{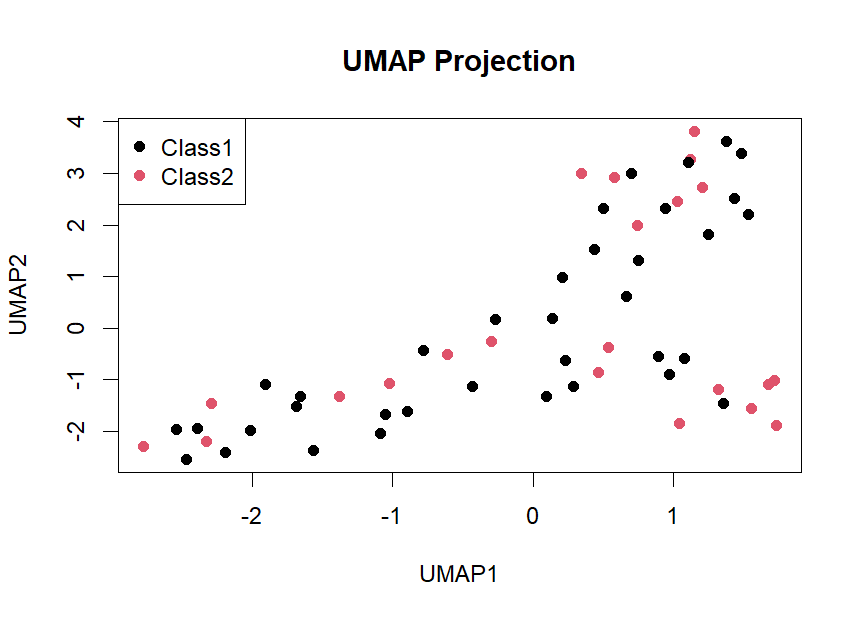}
        \caption{UMAP}
        \label{fig:umap}
    \end{subfigure}
    \caption{Embeddings of NIR spectral data using t-SNE and UMAP learning techniques.}
    \label{fig:embeddings3}
\end{figure}

To assess the effectiveness of the different algorithms, the high-dimensional NIR spectra were projected onto two- or three-dimensional spaces. The resulting embeddings for each method are visualized in Figures~\ref{fig:embeddings1}, \ref{fig:embeddings2} and \ref{fig:embeddings3}.

\begin{figure}[h!]
    \centering
    \begin{subfigure}[b]{0.45\textwidth}
        \centering
        \includegraphics[width=\textwidth]{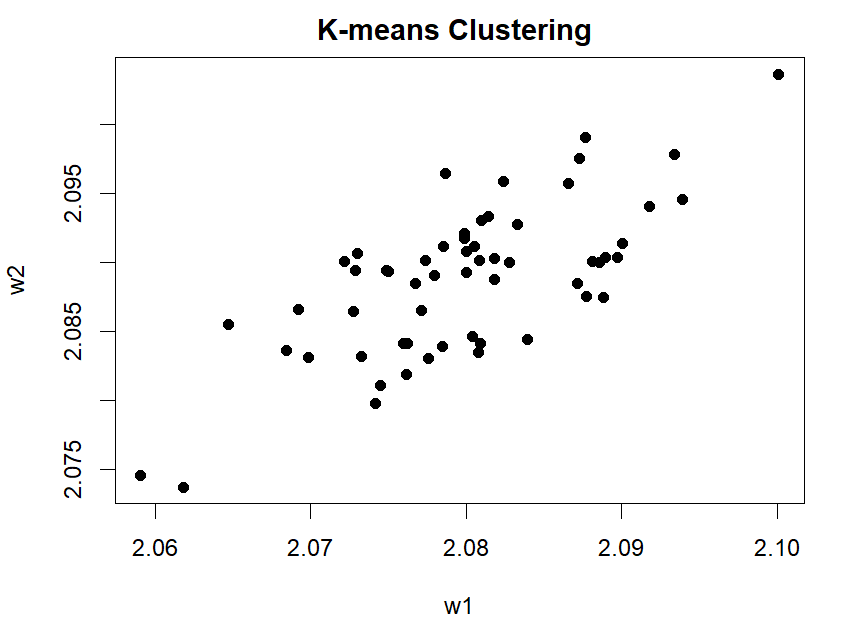}
        \caption{k-means on original data}
        \label{fig:kmeans_original}
    \end{subfigure}
    \hfill
    \begin{subfigure}[b]{0.45\textwidth}
        \centering
        \includegraphics[width=\textwidth]{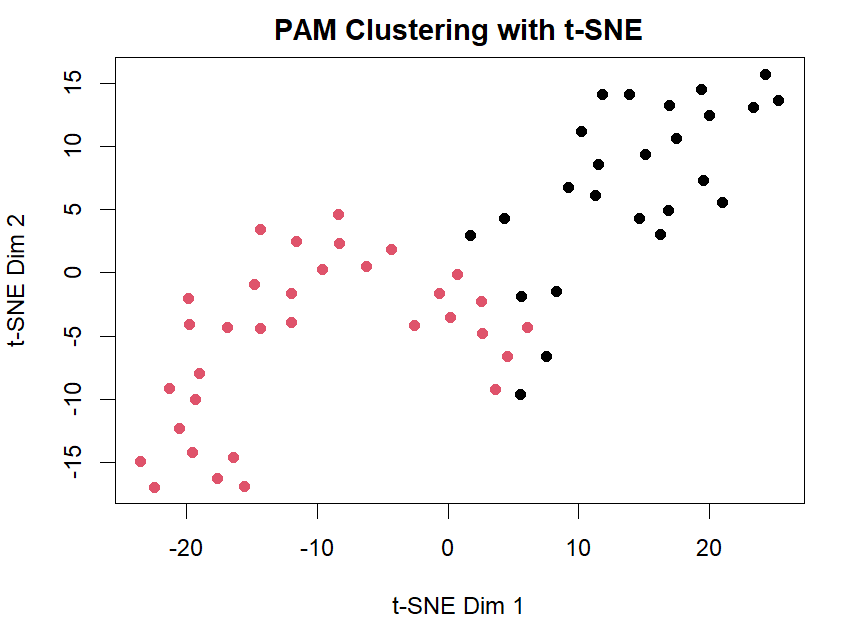}
        \caption{PAM on t-SNE embedding}
        \label{fig:pam_tsne}
    \end{subfigure}
    \caption{Comparison of clustering results: (a) k-means on the original high-dimensional NIR spectra, and (b) PAM on the t-SNE reduced embedding.}
    \label{fig:clustering_results}
\end{figure}
\vspace{2em}

Several observations can be drawn from the visualizations of these graphs:

\begin{itemize}
    \item \textbf{PCA:} As expected, PCA revealed the global variance structure in the dataset. The first two principal components captured approximately 100\% of the total variance. However, PCA failed to clearly separate samples into distinct clusters, highlighting its limitation in capturing non-linear relationships. Additionally, the clustering pattern observed in linear PCA differs noticeably from that in non-linear KPCA, suggesting the presence of non-linear structures in the dataset. This observation is consistent with the conclusions reached in our previous investigations~\cite{Sow2022, sow2022-1}.

    \item \textbf{Kernel PCA and Sparse Kernel PCA:} Both KPCA and its sparse variant provided improved separation of overlapping spectral regions compared to linear PCA. Sparse KPCA, in particular, achieved this while using fewer support vectors, offering a more interpretable and computationally efficient representation without sacrificing essential structural information.

    \item \textbf{t-SNE:} The t-SNE algorithm produced distinct and well-separated clusters, indicating that the spectral data contains meaningful groupings. It effectively preserved local neighborhood structure but showed sensitivity to parameter settings such as perplexity. Moreover, the global arrangement of clusters was less consistent, a known limitation of t-SNE.

    \item \textbf{UMAP:} UMAP demonstrated strong performance, generating compact and well-separated clusters while preserving both local and global relationships. This technique is also computationally efficient, making it particularly suitable for exploratory data analysis.
\end{itemize}

\subsection{Comparison and Interpretation}

shows a side-by-side comparison of the two-dimensional embeddings produced by each method. UMAP and t-SNE showed clear cluster separation, which may correspond to variations in chemical composition, noise level, or preprocessing differences. PCA and KPCA provided valuable global structure, but with limited cluster resolution.

\subsection{Clustering Performance}

To evaluate clustering effectiveness, k-means (formalized by Lloyd~\cite{lloyd1982lsq}) was applied directly to the original high-dimensional NIR spectra, while PAM (Partitioning Around Medoids, introduced by Kaufman and Rousseeuw~\cite{kaufman1990finding}) was applied to the lower-dimensional embedding obtained from t-SNE. The choice of applying PAM on the t-SNE embedding was motivated by the fact that the original data has more wavelengths than samples, which can negatively affect clustering performance in the high-dimensional space.

The results indicate that PAM clustering on the t-SNE reduced space produces more distinct and meaningful clusters compared to k-means applied on the original data. This highlights the benefit of using dimensionality reduction prior to clustering to capture the underlying data structure more effectively.

Figure~\ref{fig:clustering_results} illustrates the clustering outcomes for both methods, showing clearer cluster separation in the PAM + t-SNE embedding.

\subsection{Summary of Findings}

Overall, UMAP and t-SNE emerged as the most effective techniques for revealing meaningful structures in the NIR spectra of paracetamol. While PCA and KPCA provided useful variance-based insights, they were less effective in uncovering the non-linear relationships inherent in the data.

\section{Conclusions}\label{sec:conclusions}

In this study, we evaluated several dimensionality reduction techniques—namely Principal Component Analysis (PCA), Kernel PCA (KPCA), Sparse Kernel PCA, t-Distributed Stochastic Neighbor Embedding (t-SNE), and Uniform Manifold Approximation and Projection (UMAP)—applied to near-infrared (NIR) spectra of paracetamol. These unsupervised learning methods proved effective in simplifying the high-dimensional spectral data and revealing its underlying structure.

Our results indicate that while linear methods like PCA provide a fast and interpretable summary of the variance in the data, they are limited in their ability to capture non-linear relationships. In contrast, non-linear approaches such as t-SNE and UMAP more effectively uncovered meaningful clusters and local patterns within the spectra.

This work highlights the potential of integrating NIR spectroscopy with modern machine learning techniques to enhance data exploration and interpretation in pharmaceutical research. Such an approach can facilitate a deeper understanding of complex datasets, ultimately improving decision-making in quality control, formulation development, and process monitoring.

\section{Data Availability}

The NIR spectral dataset used in this study is available from the authors upon reasonable request. Any additional code or materials related to dimensionality reduction and clustering analyses can also be provided to support reproducibility and further investigation.

\section*{Acknowledgments}

Aminata Sow gratefully acknowledges financial support from the Swedish International Development Cooperation Agency (SIDA) through the International Science Program (ISP) at Uppsala University during her doctoral studies.

The authors also thank their colleagues for insightful discussions and valuable feedback throughout the course of this research.

\newpage
\printbibliography[heading=bibliography]

\end{document}